\documentclass[journal ]{new-aiaa}
\usepackage[utf8]{inputenc}
\usepackage{textcomp}
\usepackage{acro}
\usepackage{graphicx}
\usepackage{amsmath}

\usepackage[version=4]{mhchem}
\usepackage{siunitx}
\usepackage{longtable,tabularx}
\usepackage{float}
\usepackage{enumerate}
\usepackage{comment}
\usepackage{enumitem}

\title{Safety Analysis of eVTOL Operations based on STPA}

\author{Mariat James Elizebeth \footnote{Lead Engineer, WMG, University of Warwick, 6 Lord Bhattacharyya Way, Coventry, United Kingdom, CV4 7AL(Corresponding Author).}}
\author{Shufeng Chen \footnote{Lead Engineer, WMG, University of Warwick, 6 Lord Bhattacharyya Way, Coventry, United Kingdom, CV4 7AL.}}
\author{Halima El Badaoui \footnote{Project Engineer, WMG, University of Warwick, 6 Lord Bhattacharyya Way, Coventry, United Kingdom, CV4 7AL.}}
\author{Siddartha Khastgir \footnote{Professor, WMG, University of Warwick, 6 Lord Bhattacharyya Way, Coventry, United Kingdom, CV4 7AL.}}
\author{Paul Jennings\footnote{Research Director, WMG, University of Warwick, 6 Lord Bhattacharyya Way, Coventry, United Kingdom, CV4 7AL.}}
\affil{WMG, University of Warwick, 6 Lord Bhattacharyya Way, Coventry, United Kingdom, CV4 7AL.}

\begin{document}
\maketitle
\begin{abstract}
Electric Vertical Take-Off and Landing (eVTOL) aircraft are expected to be quieter and more cost-effective than helicopters, offering major economic and social benefits through improved connectivity. Their adoption will require new ground infrastructure and airspace redesign, introducing risks involving multiple stakeholders (Regulators, eVTOL operators, Air navigation service providers, Vertiport operators, OEMs, Pilots, etc.). To assess these risks for the UK airspace,  systems-thinking based System Theoretic Process Analysis (STPA) was conducted. To manage the large number of Unsafe Control
Actions (UCAs) and requirements generated due to the complexity of the
analysis, a novel extension to STPA for the prioritization of results was applied. 317 UCAs were
identified in total out of which 110 high-priority UCAs were analyzed (Step-4), resulting
in 377 causal factors and 432 requirements. These were prioritized to produce a targeted list of 124 distinct high-priority requirements, 56 of which were identified as gaps in existing aviation regulations, policies, or procedures.. These highlight opportunities for regulatory updates in areas such as organizational performance, certification processes, training, collision avoidance, energy management, and automation.
The findings provide regulators with safety considerations that could shape new or updated regulations, compliance methods, and guidance materials for the safe deployment of eVTOLs.
\end{abstract}
\section{Introduction}
\lettrine{A}{viation} and information technology
are developing rapidly. In the United States, the Federal Aviation Administration (FAA) and National Aeronautics and Space Administration(NASA) have collaboratively introduced the notion of Advanced Air Mobility (AAM). The goal is to create a new aviation transportation system employing electric aircraft for moving both passengers and cargo \cite{authority2020urban}. Over the past four decades, Unmanned
Air Vehicles (UAVs) have been primarily used in military settings for tasks such as tracking, surveillance, weapon engagement, and collecting air-borne data. However, due to their lower production and operational expenses, adaptability in design to meet customer requirements, and the elimination of pilot risk in challenging missions, UAVs have also found a strong commercial appeal \cite{sarris2001survey}. Despite their benefits, civilian UAV systems face significant challenges related to airspace integration with manned flights, certification standards, reliability, and ensuring flight safety, which must be thoroughly resolved before their widespread adoption in everyday use \cite{ozdemir2014design}.
Numerous studies have already been conducted on different facets of Urban Air Mobility (UAM), yielding substantial progress in airspace planning \cite{bauranov2021designing}, evaluating market demand \cite{long2023demand}, and designing aircraft \cite{piccinini2020numerical}.
A specific branch of AAM, known as UAM, focuses on the regulations, processes, and technologies necessary for operating air traffic for passengers and cargo in urban areas \cite{guo2024vtol}. A report on the use of electric vertical take-off and landing aircraft (eVTOL) to transport passengers within urban environment explicitly outlined the main challenges facing the implementation of eVTOL operations, one of them being the establishment of vertical take-off and landing (VTOL) infrastructure \cite{holden2016fast}. 
With the rapid development of AAM concepts, ensuring safe deployment of eVTOL aircraft in the current aviation airspace presents unique challenges. This is because of their novel technology, operational complexity, and regulatory uncertainties. These challenges span various stakeholders of the eVTOL operation, from design and airworthiness at the eVTOL manufacturer side, to the operational and infrastructure management at the vertiports or aerodromes, and the humans (passengers and the public). The UK Airspace Modernization Strategy emphasizes integration  rather than segregation of the various groups of airspace users, each of which has distinct operational needs and capabilities. The incorporation of new technologies, such as eVTOL, into the existing air traffic management system introduces new risks and hazards that need to be identified and assessed.
 Even though regulatory bodies have been crafting specific regulations for eVTOLs, the intricate interactions among various eVTOL operation stakeholders mean that the established requirements cannot yet be considered comprehensive or thoroughly evaluated. Moreover, the urgent need to implement a potentially high volume of requirements has added to the complexity of identifying and managing them. \par An operational concept at the beginning stages of its development provides both significant challenges and substantial opportunities. A concept that is not fully developed may often miss thorough safety and risk evaluations due to limited risk modeling choices. Nonetheless, conducting safety evaluations early on is beneficial as it allows for more flexibility in implementing safety-oriented design alterations, which become harder to enact later in the development process \cite{harkleroad2013review}. Continually identifying tools, techniques, and strategies is essential for simultaneous concept development and safety assessments of NextGen operational enhancements. The integration of robust hazard analysis methods into the overall systems engineering process is essential to ensure that safety is embedded into systems right from their inception. The main obstacle to achieving this goal is the current analytical tools' ineffectiveness during the initial stages of concept development \cite{fleming2015integrating}. Existing safety methods are applicable in the later stages of system development, when detailed 
design information is available and are inadequate for analyzing systems that exhibit emergent behavior, particularly those involving multiple stakeholders and interactions between humans and technology \cite{carayon2015advancing}.

\subsection{Literature Review }

\par Traditional safety analysis methods such as Fault Tree Analysis (FTA) \cite{hua2025research}\cite{sotoodeh2024safety}, Event Tree Analysis (ETA) \cite{li2023eta}, Failure Mode and Effects Analysis (FMEA) \cite{daugistanli2025failure}, Hazard and Operability Study (HAZOP) \cite{dang2025risk}, and Bow-Tie Analysis \cite{faturachman2023safety}, which are based on the Linear Chain of Events model,  have been widely used in the aviation domain for decades. These methods provide a structured approach to analyzing and mitigating risks. However, when applying the same methods to complex systems with novel technologies like eVTOLs, the unsafe interactions between the components of the system may be overlooked. Systems- Theoretic Process Analysis(STPA) is a modern safety analysis technique based
on the new accident causation model - System-Theoretic Accident
Model and Process (STAMP), which treats safety as a hierarchical control problem
rather than a failure problem \cite{saferworld}. STPA recognizes safety as an emergent
property of a complex system caused by the interaction of its components \cite{Castilho2018a}.

STPA is a structured method designed to identify ways in which complex systems might become unsafe and result in accidents. This is accomplished by modeling the control structure with the controllers and controlled processes, and then examining their interactions including the feedback loops, to comprehend how adverse outcomes could arise. STPA considers a diverse range of causal factors of the hazardous interactions, including flawed control algorithms due to flaws in their requirements, communication errors, and delays, conflicted controls, processing delays, misinterpretations of the received data or signals, etc. There exists work in various domains
that present cases where STPA identified hazards previously not identified by
the traditional analysis techniques \cite{mallya2016using}. Numerous
studies on comparison of STPA with traditional techniques such
as FTA, FMEA and HAZOP have shown that STPA found all the causes identified by
these traditional methods as well as additional causes including those
related to software and system design \cite{james2023comparison}\cite{sun2022comparison}\cite{heikkila2023hazard}\cite{bensaci2018comparative}\cite{merrett2019comparison}. A comparison of
FMEA and STPA applied to the case study of a forward collision avoidance
system concluded that both methods complemented each other as STPA
was able to find more software error type hazards while FMEA identified
more component failure type hazards than STPA \cite{sulaman2019comparison}. Studies have also
demonstrated integration of STPA with FMEA technique for risk assessment,
as STPA does not include risk evaluation \cite{La2019}\cite{chen2020novel}. An approach involving the
application of STPA in addition to FTA, to identify the potential hazardous
scenarios and their causal factors in a robotic analysis laboratory, is presented
in \cite{bensaci2020new}.
STPA has shown promising results across various industries, including space applications\cite{Ishimatsu2014a}\cite{Owens2008}, aviation \cite{Castilho2018a}\cite{liu2024safety}\cite{lu2025integrated}, medical
\cite{Balgos2012a}\cite{Silvis-Cividjian2020}, defence \cite{Pereira2006}\cite{Stanton2019}, process \cite{sultana2019hazard},rail \cite{oginni2023applying}, marine \cite{qiao2023causation}\cite{hullein2024using}\cite{li2024risk}\cite{yuzui2025toward}\cite{nakashima2025addressing} and
automotive \cite{Abdulkhaleq2018}\cite{Schmid2019}\cite{Sulaman2014a} \cite{elizebeth2025hazard}industries. The
state-of-the-art advances, future research trends and practical applications of
STPA have been illustrated in \cite{Zhang2022}.
\par There is limited research available on eVTOLs, especially concerning the application of STPA. A conflict resolution security analysis method for low altitude UAV based on STPA is 
proposed in \cite{liu2024safety}. A preliminary risk
evaluation conducted on three critical agents: the product (the eVTOL aircraft), the manufacturer
(responsible for building and certifying the eVTOL), and the operator (whether human or
autonomous) to support regulatory agencies in the certification process of eVTOL
aircraft is presented in \cite{filhopreliminary}. A safety-oriented architecture design
process of flight control system for eVTOL based on a comprehensive
method integrating ARP4761 and STPA has been proposed in \cite{ning2023safe}. Another research work identified hazards
and causal scenarios that would lead to losses in birds strikes during an eVTOL landing in urban centers using STPA\cite{de2022safety}. A report produced by the Lincoln Laboratory (on behalf of the Federal Aviation Administration) to conduct a survey of risk-based modeling and analysis
techniques to support NextGen concept assessment and validation,  recommended consideration of applying System-Theoretic Process Analysis (STPA) to NextGen
concepts early in their design to identify risk and analyze hazards \cite{harkleroad2013review}. Another report summarizing a joint effort by civil aviation authorities to evaluate STPA and its applicability to aviation safety involving Subject Matter Experts (SMEs) from various organizations such as the FAA, European Union Aviation Safety Agency (EASA), International Civil Aviation Organization(ICAO), and NASA concluded that STPA provides a capability beyond current practices to identify interactions and scenarios relevant to regulatory safety objectives that is applicable to future technologies like increasing autonomy and eVTOL \cite{thomas2024evaluation}. A technical report on STPA analysis of NextGen Interval management components found that STPA is capable of managing the advanced functionalities of NextGen systems and the complexity of the proposed operational enhancements. It offers a structured approach to support subject matter experts—essentially enabling a systematic investigation \cite{fleming2013technical}. A practical
deployment framework for the integration of eVTOL aircraft into public airspace, in accordance with
the FAA special airworthiness criteria is presented in \cite{bridgelall2023integrating}. An application of the Functional Resonance Analysis Method (FRAM) in
assessing the operational concept of eVTOLs in
urban environments highlighted the importance of integrating diverse stakeholders and
addressing uncertainties in urban air mobility planning\cite{fowlerfram}.

\par
To assess the risk posed by eVTOL operations in UK’s current airspace, a systems thinking based safety analysis method- STPA , was chosen. To ensure the effectiveness of STPA, cooperation between two types of experts was necessary: (1) STPA specialists and (2) domain experts. The authors brought their STPA expertise, while various stakeholders participated to provide their perspectives on eVTOL operations. These stakeholders encompassed the national aviation regulator, eVTOL operators , Air Navigation Service providers,  Vertiport Operators, Original Equipment Manufacturers , as well as Helicopter pilots.

\subsection{Research Objectives and Contribution}
This paper offers a distinctive contribution to research in the field of advanced air mobility, by presenting STPA analysis of eVTOL operations between London Heliport and Silverstone Aerodrome in the UK, as an example. This example represented the typical high-frequency daily utilization expected for standard eVTOL operations. The results would however be applicable to eVTOL operations between any two licensed Vertiports/Aerodromes. The objective of this research was to investigate the emergent behaviors associated with integration of a novel technology like eVTOL into the current airspace. This is due to the fact that the existing airspace design and regulations framework may not be suitably equipped to handle these unexpected behaviors safely. The analysis was split into an organizational analysis and operational analysis. Organizational analysis primarily focussed on modeling and analyzing the interactions between the organizations prior to the flight operations. Operational analysis focussed on the interactions for specific flight operations on the day. The result is a set of recommendations to be developed into practical, implementable actions for each relevant stakeholder in the form of regulatory and policy updates, operational instructions, and practical guidance. This paper presents the first comprehensive application of STPA to eVTOL Operations in the United Kingdom.

STPA can identify numerous UCAs and requirements depending on the level of granularity of the analysis and the complexity of the system being analyzed. Managing these large number of results can become challenging. Consequently, there has been extensive research on methods to enhance the STPA process to manage the large number of UCAs and requirements, especially when dealing with large, complex systems. Such an enhancement allow stakeholders and analysts to concentrate on the most crucial elements to enhance system safety. Assigning a rank to the UCAs can provide valuable insight into which are the most crucial. This enables analysts to prioritize and manage the STPA Step-4 analysis to focus on the highest-priority UCAs that require immediate mitigation, rather than attempting to address every UCA identified in Step-3, particularly when there are time and resource constraints. The STPA Step-4 analysis has the potential to produce a multitude of requirements, numbering in the hundreds, with some being redundant. A requirement prioritization framework would help the stakeholders to prioritize unique requirements and address the most critical ones, to begin with. Determining the prioritization of requirements when resources are constrained is a genuine project management challenge encountered by several organizations. It is important to highlight that the aim is not to overlook certain requirements, but rather to methodically and logically decide their order of implementation. An extension to the standard STPA methodology was created and implemented for prioritizing the results in this case study. While this paper primarily concentrates on application of STPA to eVTOL Operations which necessitated prioritization of results due to their large volume, the specific methodology used for prioritization is beyond the scope of this paper and is detailed in  separate publications \cite{elbadaoui2025structured}\cite{chen2025scalable}.
\par This paper is organized in five sections as follows: 
Section \uppercase\expandafter{\romannumeral 2}  presents the standard STPA methodology and gives an overview of the prioritization concept, developed as part of this case study. Results based on the application of STPA to eVTOL Operations is presented in Section \uppercase\expandafter{\romannumeral 3}. Discussion is presented in Section \uppercase\expandafter{\romannumeral 4} and Section \uppercase\expandafter{\romannumeral 5} concludes the paper with some future work.

\section{Methodology}
\begin{figure}[ht]
    \centering
     \includegraphics[scale=0.04]{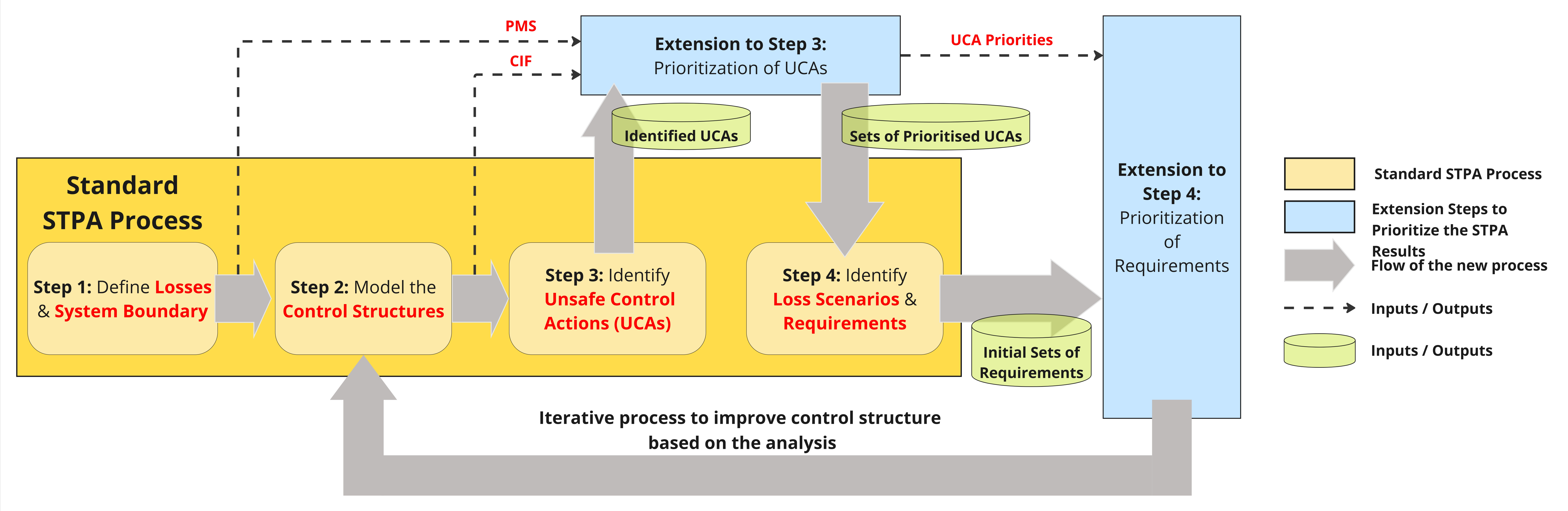}
    \caption{Standard STPA Process and extensions to Steps 3 and 4 for Prioritization }
    \label{fig:1}
\end{figure}

STAMP based STPA treats accidents as a control problem instead of a failure problem and prevents accidents by enforcing constraints on the behavior of the system. The conventional STPA approach involves four distinct steps. To handle the significant complexity and volume of results from the analysis, the authors developed a concept designed to prioritize the STPA outcomes effectively. Fig.\ref{fig:1} illustrates the standard STPA process (yellow blocks) as well as the extensions for the prioritization of STPA results (blue blocks), developed and implemented as part of this case study.

\subsection{Standard STPA methodology}
The standard STPA methodology, which is a four-step process, is described in this section. The main steps of STPA are as follows: 
\subsubsection{Step 1: Define purpose of the analysis}
STPA starts by identifying any losses which are unacceptable to the stakeholders, and system-level hazards. The system boundary is also defined in this step. The system boundary defines the range of ownership and analysis – i.e., the components outside the system boundary are not accessible to the designer for any potential upgrades required, and the components within the boundary can be appropriately managed or redesigned if the analysis outcome suggests.
\subsubsection{Step 2: Model the control structures}
The subsequent step involves developing a system model referred to as a control structure. The control structure consists of hierarchical functional blocks illustrating the functional interactions between the system components by representing the system as a series of feedback control loops. Each control loop consists of a controller that provides control actions (CA) to control some process and enforce constraints on the behavior of the controlled process. The control algorithm represents the controller’s decision- making process, it determines which control actions to provide and when. Controllers also have process models that represent the controller’s internal beliefs (which may include beliefs about the process being controlled or other relevant aspects of the system or the environment), used to make decisions. 
\subsubsection{Step 3: Identify Unsafe Control Actions (UCA)}
After identifying Control Actions in the control structure, each CA is further analyzed to identify how the CA would manifest into a UCA. Depending on the context of providing a CA, it could lead to one or multiple system-level hazards, which in turn could lead to the losses (identified in Step 1). If a CA were always unsafe, then it would never be included in the system design. The CA is analyzed using certain guide words to identify UCAs, as follows:
\begin{itemize}
    \item Not providing the CA leads to a hazard.
    \item  Providing the CA incorrectly or when not needed 
leads to a hazard.
\item  Providing a CA too early or too late or in the wrong 
order leads to a hazard.
\item  Providing the CA too long or stopping the CA too soon leads to a hazard.
\end{itemize}
In this case study, unsafe control actions (UCA) were directly associated with losses rather than hazards, which is a departure from the standard STPA methodology. This alternative approach was selected because the analysis focused on the interactions among different organizations and the eVTOL aircraft, rather than on a specific technical system (such as the aircraft itself, for example). Additionally, this method made it easier for SMEs to grasp the potential consequences of UCAs, even if they were not familiar with STPA.
\newline UCA IDs were structured as follows- UCA (Ph U)-X.Y. Z where:
\begin{itemize}
    \item U refers to the phase in which the UCA was identified. 
    \item X denoted the number of the CA.
    \item Y represented the type of UCA
    \begin{itemize}
        \item Type 1: The CA is not provided
        \item Type 2: The CA is provided incorrectly
        \item Type 3: The CA is provided when not needed
        \item Type 4: The CA is provided too early
        \item Type 5: The CA is provided too late
        \item Type 6: The CA is provided too long
        \item Type 7: The CA is provided too short

    \end{itemize}
    \item Z denoted the number of the UCA identified for the CA and type.
\end{itemize} 
 
For example, the UCA(Ph1)-18.2.1 in TABLE \ref{tab:TABLE 2} is the first UCA (i.e., Z = 1) of the CA number X = 18 (i.e., ‘Onward Clearance’) from Phase 1 (i.e. U = 1) Control Structure with type 2 - i.e., Y = 2 (The CA is provided incorrectly). 

\subsubsection{Step 4: Identify Loss Scenarios}
Once the UCAs are identified for all the control actions in the control structure, possible loss scenarios, which describe the Causal Factors (CFs) that can lead to the UCAs, are identified by analyzing the specific control loops of the Control Actions. For a UCA to occur, the process model of the controller has a belief based on which it believes that the CA it is directing is safe when it is unsafe. The causes of such beliefs can be identified based on two types of loss Scenarios –‘Type-A’ and ‘Type-B’. Type-A loss scenarios mainly explain what triggers the CAs to be unsafe. Type-B loss scenarios explain how correct CAs are not executed or are improperly executed, leading to UCAs. Once the loss scenarios are analyzed, requirements are then defined to prevent or mitigate the CFs.
\par It is worth highlighting that the STPA experts collaborated closely with domain specialists, who reviewed the findings during both in-person workshops and virtual meetings throughout the course of the analysis. These domain experts represented a wide spectrum of stakeholders, including the UK Aviation Regulator, eVTOL operators, Air Navigation Service Providers, the British Helicopter Association, Vertiport Operators, Original Equipment Manufacturers (OEMs), and helicopter pilots. 
\subsection{Extension of Standard STPA Process for Prioritization of Results}
This section gives an overview of the concept for the prioritization of the STPA Results (UCAs identified in Step-3 and requirements identified in Step-4). 
\subsubsection{Extension to Step 3: Prioritization of UCAs}
The methodology for the prioritization of UCAs is primarly based on three factors: Pre-Mitigation Severity (PMS), Controller Impact Factor (CIF), and Expert Judgment (EJ). Based on the STPA results, the initial two factors are given specific values by STPA experts, whereas the final factor, as its name suggests, relies on inputs from SMEs.
PMS is assigned based on the severity of the ranked losses (identified in STPA Step-1) that the UCAs lead to. CIF is assigned by identifying the position of the UCA Controller in the hierarchical control structure (created in STPA- Step 2). The EJ score is calculated based on values assigned by SMEs for five factors : Operational Disruption, Criticality, Detectability, Effect on Other Stakeholders, and the Likelihood of Occurrence. A UCA Prioritization Matrix which depicted criticality across five levels (from very low to very high), was created based on PMS, CIF and EJ. The prioritization of UCAs concept and the creation of UCA Prioritization matrix is detailed in \cite{elbadaoui2025structured}.
\subsubsection{Extension to Step 3: Prioritization of Requirements}
The methodology for prioritization of requirements developed as an extension to the standard STPA methodology is briefly described in this section. This was applied to the requirements from the Step-4 analysis(identification of loss scenarios) of the high-priority UCAs identified using the application of the Prioritization of UCAs concept mentioned in the section above.
There are two key factors considered for the prioritization of Requirements: 1) UCA Priority; and 2) Requirement Score.
The UCA Priority is defined as the product of EJ and CIF that were identified in the UCA Prioritization step. The Requirement Score of each requirement is calculated based on four factors for which values are again assigned by the experts: Time; Cost; Type of requirement; and Likelihood of occurrence. The concept for prioritization of requirements and generation of the Requirement prioritization matrix is detailed in \cite{chen2025scalable}.
\par Following the application of the  prioritization of the STPA Results described above, the SMEs conducted further analysis of the high-priority requirements to determine whether they were gaps within the current aviation regulations, policies, or procedures. This gap analysis elicited a significant number of potential issues for the regulator and industry to assess and consider, as detailed in Section \uppercase\expandafter{\romannumeral 3}.
\section{Results}
The results of the application of STPA to eVTOL Operations, using the methodology described in Section \uppercase\expandafter{\romannumeral 2}, is detailed here. 
The assumptions made for the analysis are listed below.
\par Assumptions about eVTOL features/ characteristics:
\begin{itemize}
    
  \item Electrically powered 
    \item  Seating capacity in the range of 4 to 6 passengers with one onboard pilot
    \item  No autonomous flight capabilities 
\end{itemize}
 \par Assumptions about the Operating Environment:
\begin{itemize}

    \item A service area focused on a large modern city with features including an urban metropolitan landscape, high-rise buildings or a major airport hub.
    \item Airspace shared with other manned and unmanned air traffic and eVTOLs.
    \item Operation under Visual Flight Rules (VFR); in Visual Meteorological Conditions (VMC). 
    \item Infrastructure sufficient for battery charging, dispatch, passenger management, and other associated needs, including purpose-built vertiports for take-off and landing are available. 
\end{itemize}		
The eVTOL aircraft itself was treated as a black box and its design was not analyzed for potential causes that can lead to accidents.

\subsection{Step 1: Define Purpose of the Analysis}
As part of STPA Step 1, the losses that were unacceptable to the stakeholders were identified. An initial list of losses was defined by the STPA analysts based on prior experience. These losses were then reviewed and ranked by the stakeholders, based on their stake in the system, i.e. what they valued and what their goals were.
TABLE \ref{tab:TABLE 1} shows the list of ranked losses. Safety-critical losses (example – L1) were ranked high while non-safety critical or business losses (example- L-4, L5) were ranked low.
\begin{table}[hbt!]
\caption{\label{tab:TABLE 1} List of Losses }
\centering
\begin{tabular}{clc}
\hline
ID& Losses&Ranking\\\hline
 L1 & Loss of life or injury to 1st, 2nd or 3rd parties &1  \\
 L2 & Loss of or damage to the eVTOL or surrounding item/property/infrastructure &2  \\
 L3 & Loss of transportation mission &3  \\
 L4 & Loss of customer satisfaction or public confidence in eVTOL&4  \\
 L5 & Loss of business goal of eVTOL Operator&5  \\
 \hline
\end{tabular}
\end{table}
\par \textit{NOTE:}  For L-1, 1st party refers to the eVTOL crew, 2nd party refers to passengers and 3rd party refers to any one external to the eVTOL aircraft.
\par Based on discussions with the stakeholders, the system boundary was also defined in this step. The system under analysis comprised of the Regulator (UK CAA), the Local Authorities (associated with the take-off and landing sites), Air Navigation Service Provider (NATS), Licensed Vertiports/Aerodromes (at take-off and landing sites), the eVTOL Operator, eVTOL aircraft (including the Commander) and the eVTOL Manufacturer. The local landowners (at both take-off and landing sites), the local emergency authorities and the infrastructure providers - UK Power Network Operator and UK Data Network Operator, were excluded from the system (outside the system boundary).
\subsection{Step 2: Model the Control Structures}

\begin{figure}[ht]
    \centering
    \includegraphics[scale=0.5]{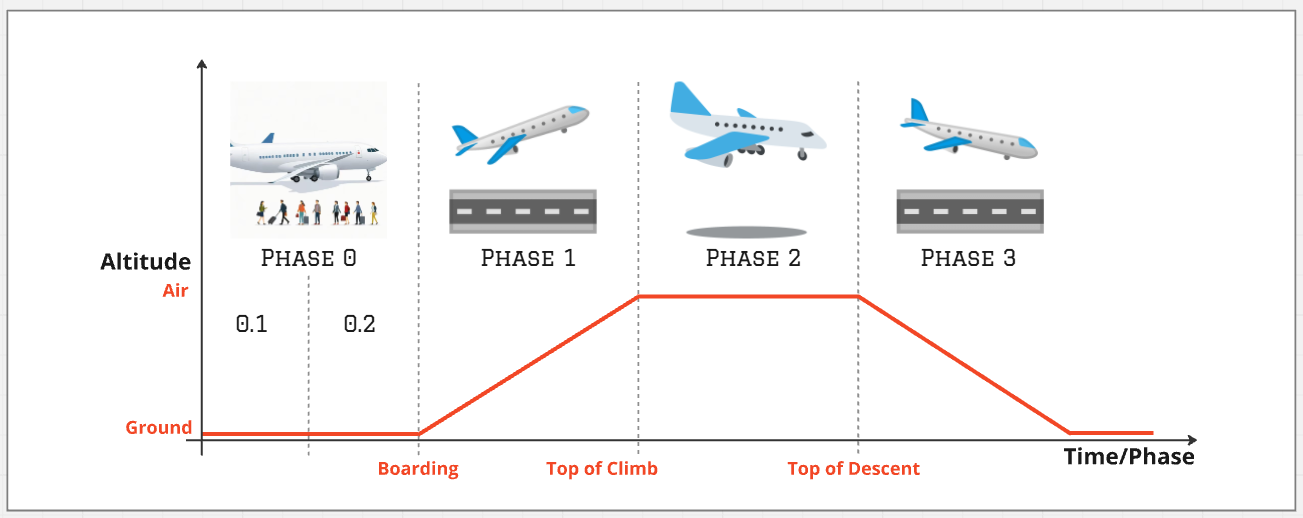}
    \caption{Flight Phases for eVTOL Operations}
    \label{fig:2}
\end{figure}

The analysis was split into an organizational analysis and operational analysis. Organizational analysis primarily focused on modeling and analyzing the interactions between the organizations prior to the flight operations. Operational analysis focused on the interactions for specific flight operations on the day. The analysis was split into five phases - Phase 0.1, Phase 0.2, Phase 1, Phase 2 and Phase 3, as illustrated in Fig.\ref{fig:2}. Distinct control structures were developed for each phase, resulting in a total of five control structures. This was done to make the complex analysis more manageable and for better readability of the control structures. Phase 0 was split into two phases – a) Phase 0.1- which focused on organizational coordination, including regulatory preparations for the flight and b) Phase 0.2- operational coordination leading up to the flight(including passenger boarding). The five control structures were created at the highest level of abstraction - Level 1, using the tool Astah System Safety. A Level 1 control structure identifies the primary components and how they interact, whereas a Level 2 control structure models detailed decompositions of these primary components and interactions at the level of sub-components.

This section defines the five phases of the analysis and presents the corresponding control structures. The five phases were as follows: 
\subsubsection{Phase 0.1- Regulatory Preparation}
\begin{figure*}[p]
    \centering
   \includegraphics[angle=90, width=0.65\textwidth]{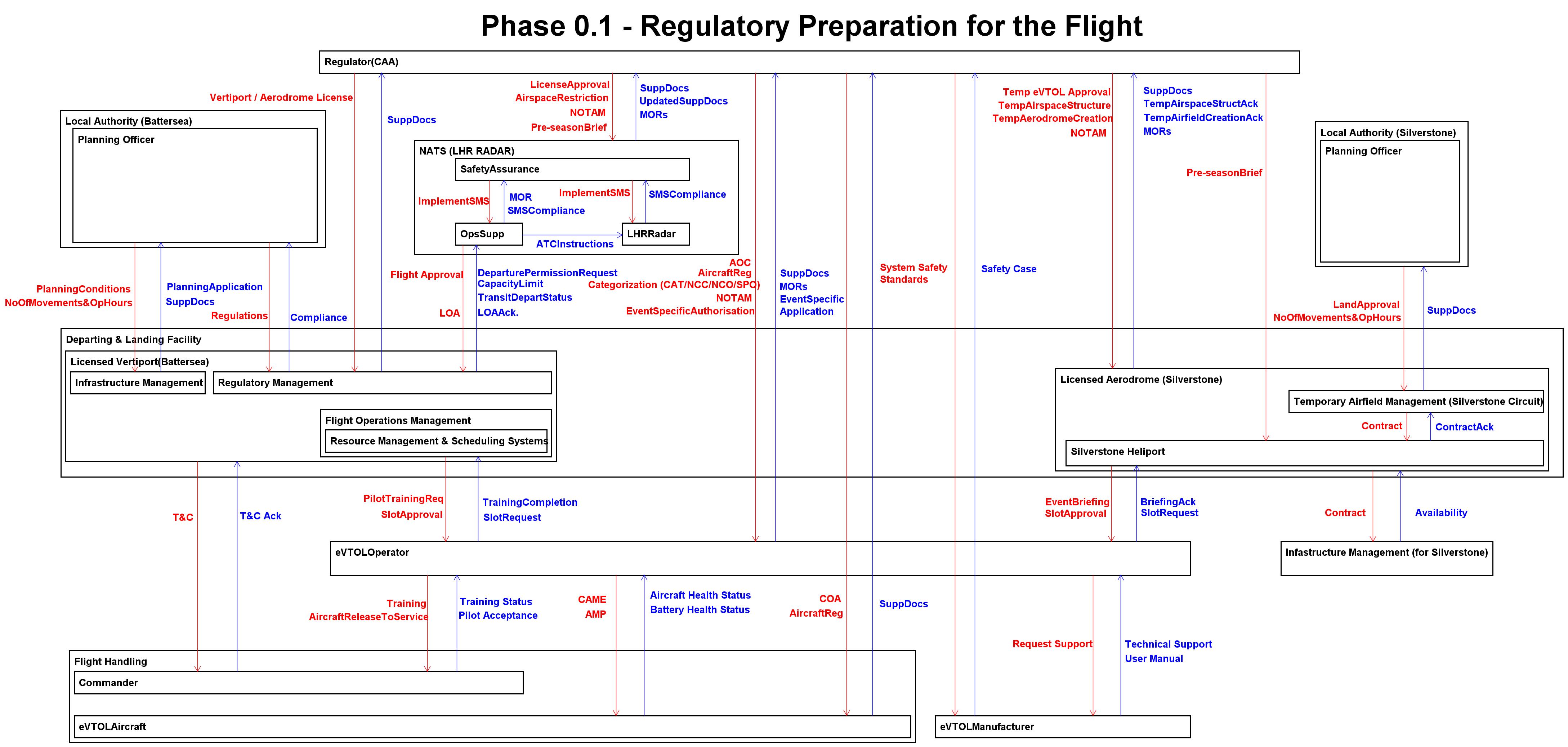}
  \caption{Control Structure for Phase 0.1}
    \label{fig:3}
\end{figure*}

Phase 0.1 covers the regulatory preparation for the flight (organizational). Fig.\ref{fig:3}) shows the interactions between the various stakeholders prior to the start of flight operations. It is worth noting that the ‘Regulator’ is included only in this control structure.
\subsubsection{Phase 0.2- Operational Preparation}
\begin{figure}[H]
 \centering
 \includegraphics[angle=90, width=0.8\textwidth]{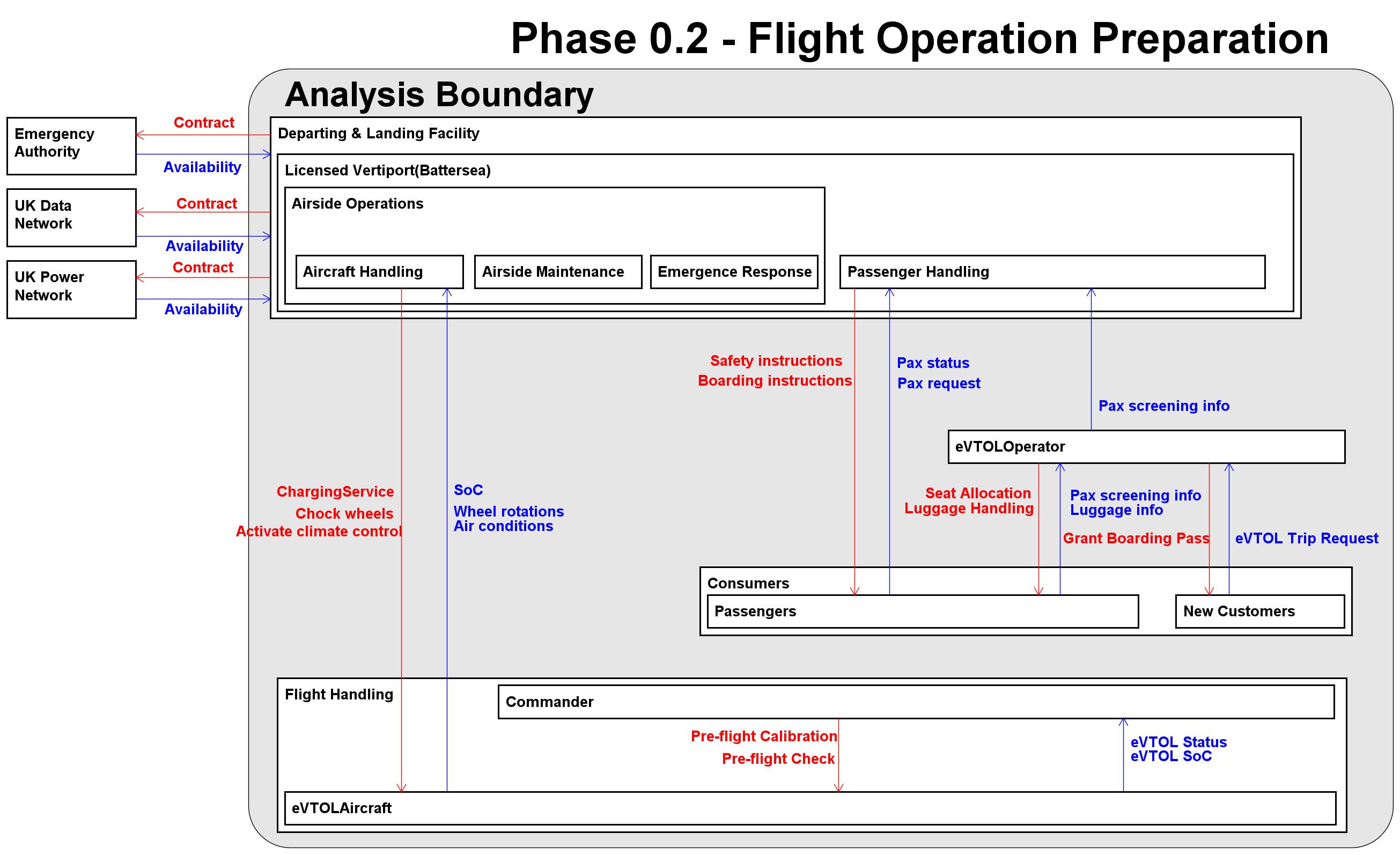}
  \caption{Control Structure for Phase 0.2}
 \label{fig:4}
\end{figure}

Phase 0.2 covers the operational preparation for the flight (until and including passenger boarding). Fig.\ref{fig:4} shows the interactions between the various stakeholders as part of the preparation for the flight operation (up to and including passengers boarding the eVTOL aircraft).
\subsubsection{Phase 1- Take-Off}
\begin{figure}[H]
  \centering
  \includegraphics[angle=90, width=0.65\textwidth]{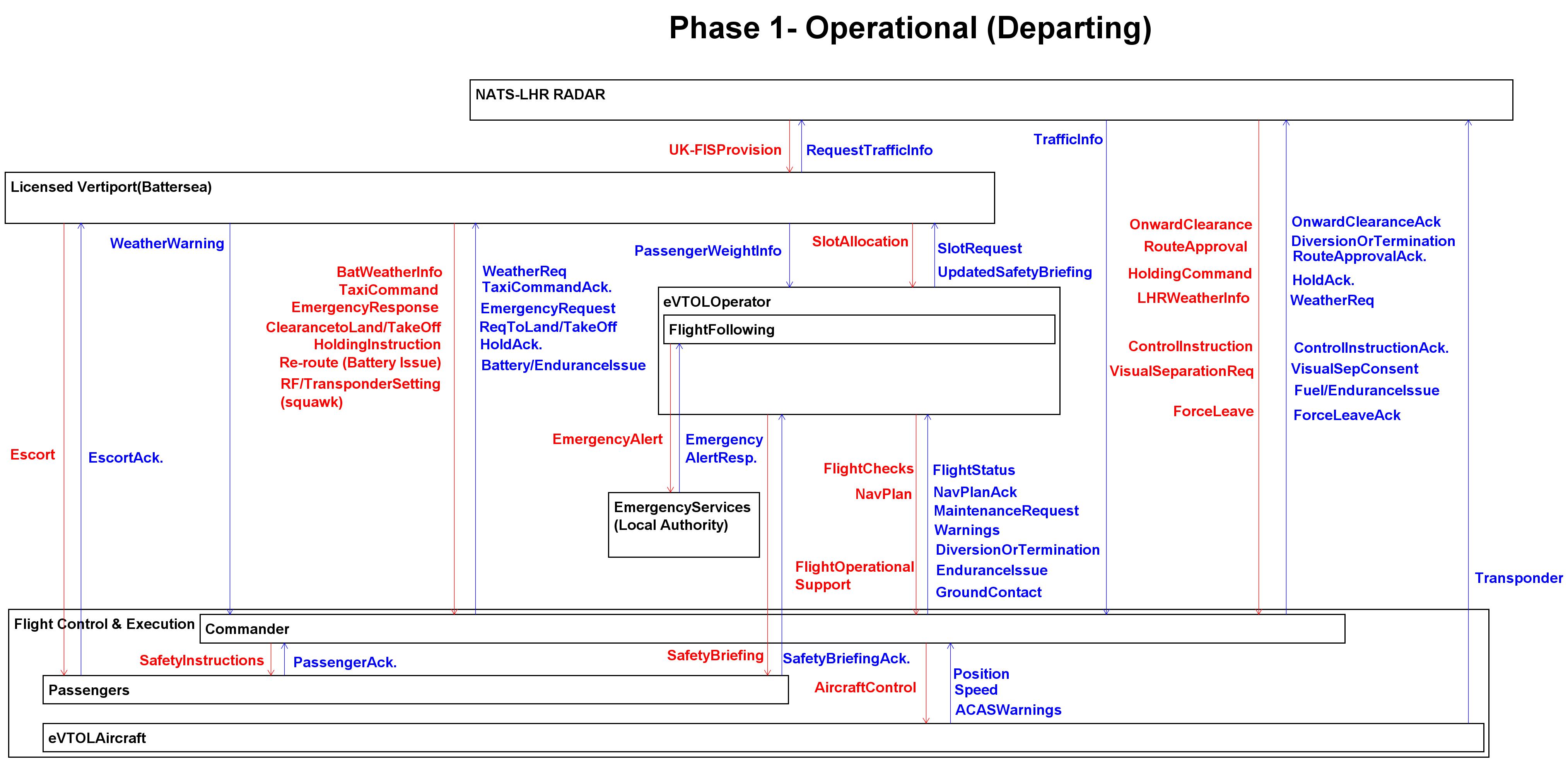}
  \caption{Control Structure for Phase 1}
  \label{fig:5}
\end{figure}

Phase 1 covers the take-off from London Heliport, when the eVTOL aircraft is in controlled airspace. Fig.\ref{fig:5} shows the interactions between the various stakeholders during the flight take-off phase (in controlled airspace).
\subsubsection{Phase 2- Cruise}
\begin{figure}[H]
  \centering
  \includegraphics[angle=90, width=0.43\textwidth]{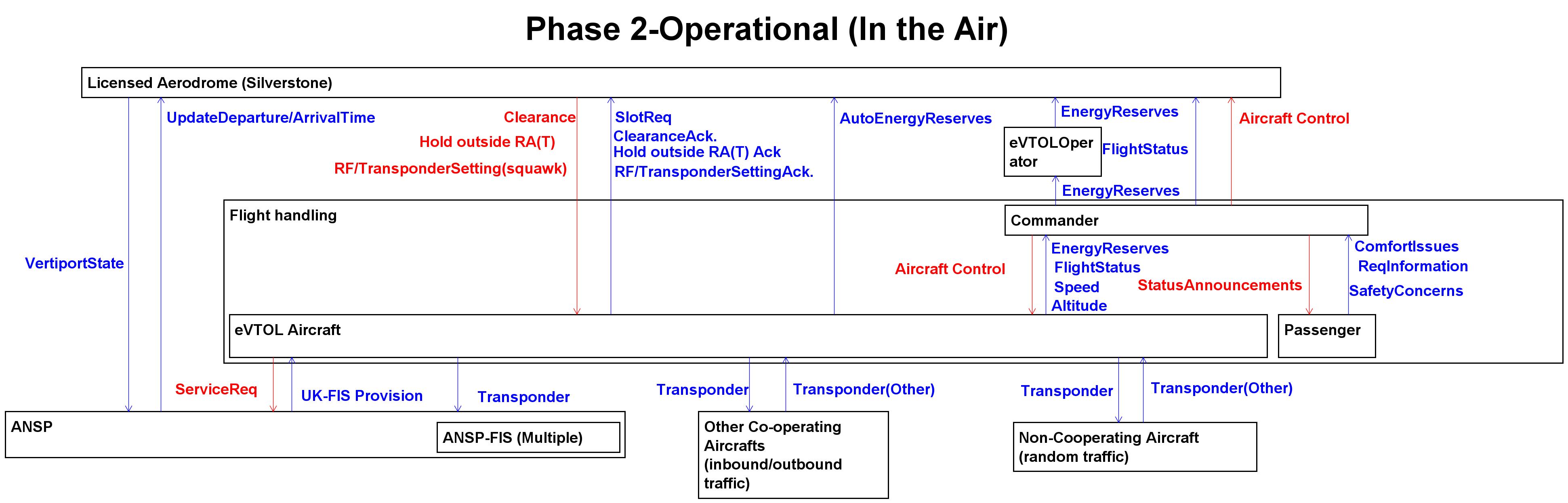}
  \caption{Control Structure for Phase 2}
  \label{fig:6}
\end{figure}

Phase 2 covers the cruise phase – i.e., after the aircraft has climbed, mostly in uncontrolled airspace. Fig.\ref{fig:6} shows the interactions between the various stakeholders during flight operation after the aircraft has climbed (major part of the trip is in uncontrolled airspace).
\subsubsection{Phase 3- Descent and Landing}
\begin{figure}[H]
  \centering
  \includegraphics[angle=90, width=0.56\textwidth]{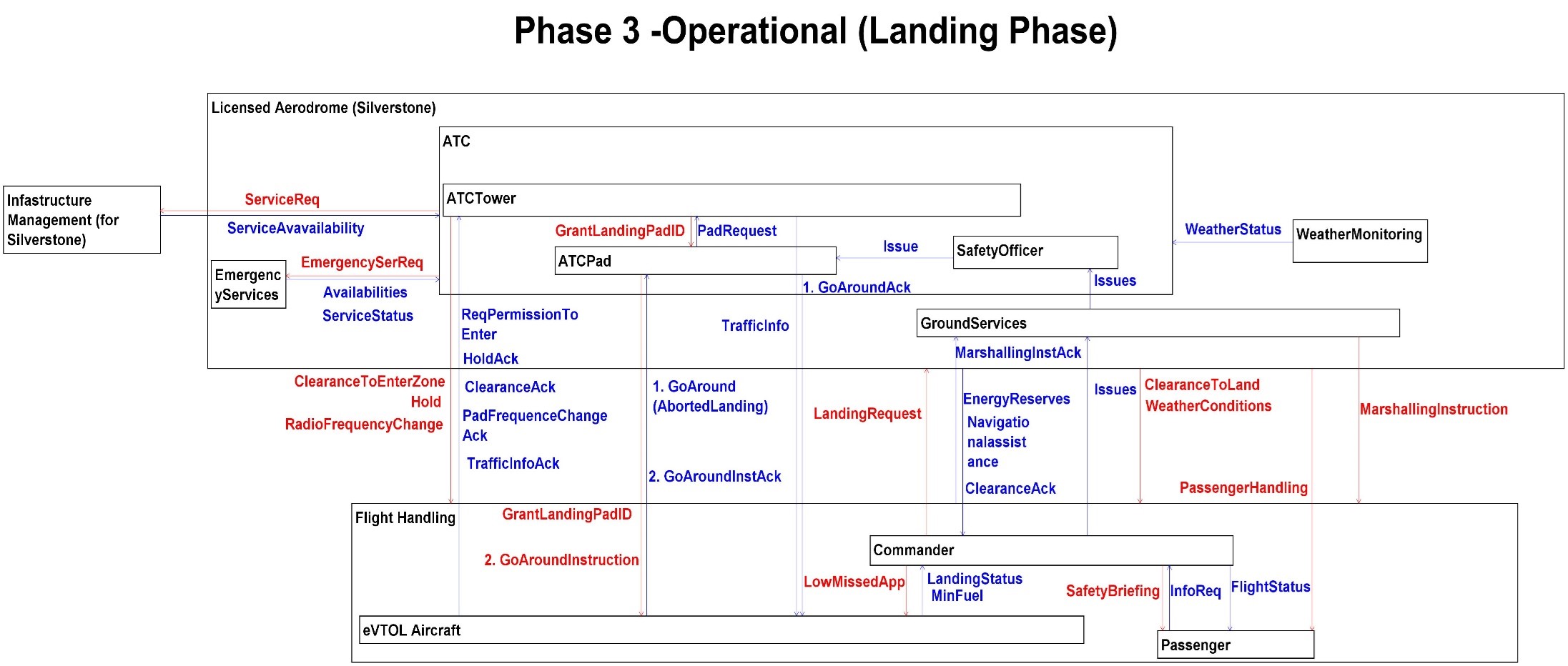}
  \caption{Control Structure for Phase 3}
  \label{fig:7}
\end{figure}

Phase 3 covers the flight operation from the start of the descent to completion of landing (landing at Silverstone Aerodrome). Fig.\ref{fig:7} shows the interactions between the various stakeholders during this flight landing phase. 

\subsection{Step 3: Identify Unsafe Control Actions}
After the creation of control structures, UCAs were identified for every CA in the control structures, as part of STPA Step-3. There were 317 UCAs identified in total across the five phases . As mentioned in Section  \uppercase\expandafter{\romannumeral 2} , for this case study, the UCAs were directly linked to losses instead of hazards, deviating from the typical STPA process. As the number of UCAs identified was high, it was necessary to prioritize the UCAs to streamline the activities for the next phase (identification of loss scenarios and requirements) and focus on the most safety-critical UCAs (which led to the highest ranked losses identified in STPA-Step-1). Based on the application of the prioritization concept detailed in \cite{elbadaoui2025structured}, the UCAs were populated in a UCA Prioritization Matrix and assigned priorities from P1 (Highest) to P5 (Lowest). 

\par The UCAs linked to each stakeholder were grouped, reviewed and updated. A sub-set of the STPA step-3 results showing some of the UCAs linked to different stakeholders- Regulator, NATS, eVTOL Operator, Licensed Vertiport/Aerodrome and Commander, in different phases is presented in TABLE \ref{tab:TABLE 2}.

Following the application of the concept for the prioritization of UCAs detailed in \cite{elbadaoui2025structured},110 High priority (those assigned P1 \& P2) UCAs were taken up for the identification of loss scenarios and requirements in Step-4.

\begin{table}[hbt!]
\caption{\label{tab:TABLE 2} Some of the UCAs identified as part of STPA Step-3}
\centering
\begin{tabular}{p{0.20\linewidth} p{0.60\linewidth} }
\hline
 \\
 \textbf{UCA-ID } & \textbf{UCA Description} \\
 \hline
UCA(Ph0.1)-28.2.1 & \textbf{Regulator} reissues ‘Vertiport / Aerodrome Licence’ incorrectly (e.g., with insufficient risk assessments) when the vertiport is actively being used for flight operations [Leading to L-1,2]  \\
UCA(Ph0.1)-24.5.1 & \textbf{Regulator} Regulator provides categorization too late when the flight is being planned.
\textit{Note:} this would delay the progress of the planning, leading to business-critical losses. [Leading to L-3,4,5]  \\
 UCA(Ph1)- 18.2.1 & \textbf{NATS (LHR RADAR)} provides ‘Onward Clearance’ incorrectly (incorrect height, routing) when there is a conflict (proximity to other aircraft, such as eVTOLs, helicopters, and fixed wing) [Leading to L-1,2,3,4,5]   \\
 UCA(Ph1)- 20.7.1 & \textbf{NATS (LHR RADAR)} stops providing HoldingCommand too soon when  the eVTOL aircraft is in hold and conflict prevails. [Leading to L-1,2,3]   \\
 
 UCA(Ph0.1)-50.2.1 & \textbf{eVTOL Operator} provides ‘Aircraft Release To Service’ for aircraft despatch incorrectly when adequate checks on the aircraft have not been carried out, this has not been detected, and the eVTOL aircraft flies [Leading to L-1,2]   \\
 UCA(Ph1)- 17.1.1 & \textbf{eVTOL Operator} does not provide ‘Safety Briefing’ (on fastening seat belts, stowing away cargo and use of portable electronic devices (PED) and the eVTOL experiences a turbulence while flying [Leading to L-1,2] \\
  UCA(Ph0.1)-32.5.1 & \textbf{Licensed Aerodrome (Silverstone Aerodrome)} provides SlotApproval too late (by x weeks) when the slot request has been submitted and flight is scheduled[Leading to L-3,4,5] \\
  UCA(Ph3)- 13.2.1 & \textbf{GroundServices} provides incorrect  MarshallingInstruction when the eVTOL  is in the landing phase at a vertiport with low visibility conditions[Leading to L-1,2,3]  \\
 UCA(Ph1)- 25.5.1 & \textbf{Commander} provides ‘Aircraft Control’ too late when the eVTOL aircraft is about to collide with an object (e.g.: infrastructure or other aircraft/drones)[Leading to L-1,2]  \\
 UCA(Ph1)-15.1.1 & \textbf{Commander} The Commander does not provide SafetyInstructions(on location of emergency exits)  ,the eVTOL aircraft flies and an emergency occurs requiring the passengers to disembark the flight swiftly[Leading to L-1]  \\
\hline
\end{tabular}
\end{table}

\subsection{Step 4: Identify Loss Scenarios }
In this step, all the high priority UCAs from the previous step (Step-3) were analysed to identify the potential CFs that could lead to the occurrence of these UCAs. Later, requirements were proposed to prevent or mitigate these CFs. The Step-4 analysis of 110 high-priority UCAs resulted in the identification of 377 CFs. 432 requirements were proposed to prevent or mitigate these CFs. 
\par To deal with the large number of requirements generated, they were grouped according to stakeholders and then prioritized using the concept for the prioritization of requirements detailed in \cite{chen2025scalable}. The requirements were analyzed further to filter out all duplicates (as one requirement could address multiple CF). This yielded 124 distinct high-priority requirements being allocated to various stakeholders (Regulator – 58, Vertiport – 40, Operator – 16, and NATS – 17).
An extract of the Step-4 results showing the requirements proposed to address the various types of CFs (organizational issues, communication errors, missing Feedback/information, inadequate Control Algorithms, delayed Feedback/information etc.)   that could lead to UCAs associated with different stakeholders- Regulator, NATS, eVTOL Operator and Licensed Vertiport/Aerodrome, is presented in  TABLE \ref{tab:TABLE 3}.
\begin{table}[hbt!]
\caption{\label{tab:TABLE 3} Some of the causal factors and requirements identified by STPA Step-4}
\centering
\begin{tabular}{p{0.25\linewidth} p{0.3\linewidth} p{0.23\linewidth} } 
\hline
\textbf{UCA Description} & \textbf{Causal Factors} & \textbf{Requirements} \\
\hline
\textbf{UCA(Ph0.1)-13.2.2:} \textbf{Regulator} does not issue ‘Vertiport / Aerodrome Licence’’ when the vertiport is actively being used for flight operations.
Note: this would affect the flight operations schedule, leading to business-critical losses.
&
The supplementary documents (compliance with regulatory standards, safety management systems, training and competency of personnel, operational readiness, data integrity and cybersecurity, environmental compliance etc.) was incomplete although the licensed vertiport met the criteria to be granted vertiport / aerodrome licence approval. As a result, the licence approval was not granted.
&
\textbf{UCA(Ph0.1)-13.2.2-RQ3:} Licensed Vertiport shall ensure that the provided supplementary documents for vertiport / aerodrome licence application are complete and up to date.

\\
\textbf{UCA(Ph1)-18.2.1:} \textbf{NATS (LHR RADAR) } provides ‘Onward Clearance’ incorrectly (incorrect height, routing) when there is a conflict (proximity to other eVTOLs, helicopters and traditional aircraft)
 &
 There is an aircraft which has deviated from its flight plan and both NATS and the eVTOL Crew are unaware.
 & 
\textbf{UCA(Ph1)-18.2.1-RQ9:} There must be a mechanism for NATS to monitor and issue alerts when the performance (position, altitude, airspeed) of aircraft in flow is not within the expected values.

\\
\textbf{UCA(Ph0.1)-50.2.1:} \textbf{eVTOL Operator} provides ‘Aircraft Release To Service’ for aircraft despatch incorrectly when adequate checks on the aircraft have not been carried out, this has not been detected, and the eVTOL aircraft flies.
 &
 eVTOL Operator is unable to correctly provide Aircraft Release To Service (although it should) due to the degradation of the internal process over time (e.g., overloaded tasks, flawed process).
 & 
\textbf{UCA(Ph0.1)-50.2.1-RQ7:} Performance review of the relevant team issuing Aircraft Release To Service within the eVTOL Operator shall be conducted periodically to ensure that the team operates properly and safely.

\\
\textbf{UCA(Ph2)-6.5.1:} \textbf{Licensed Aerodrome (Silverstone Aerodrome) } provides ‘Hold outside RA(T)’ too late when airspace congestion has already built up.
 &
 The Feedback about the current state of airspace congestion is delayed.
 & 
\textbf{UCA(Ph2)-6.5.1-RQ.2:} Licensed Aerodrome (Silverstone Aerodrome) shall conduct automated self-checks of feedback systems every x sec (to be confirmed).
\\
\textbf{UCA(Ph2)-7.1.3:} \textbf{Licensed Aerodrome (Silverstone Aerodrome) } does not provide RF/TransponderSetting(squawk)  when the airspace is congested.
 &
 The aerodrome controller misinterprets the airspace data due to its unclear format.
 &
\textbf{UCA(Ph2)-7.1.3-RQ.3:} Feedback systems must standardize data presentation using visual indicators.

\end{tabular}
\end{table}

\subsection{Gap Analysis based on the STPA Results }

It is presumed that the majority of current aviation regulations will extend to eVTOL operations. Nonetheless, to substantiate this presumption, an analysis of the determined requirements against existing aviation regulations, policies and procedures was undertaken. The term \textit{‘Gap'} was used to denote a requirement proposed by STPA that is not covered by the existing regulations, policies, or procedures related to Helicopters and/or eVTOLs.

Once STPA Step-4 was finished, an evaluation was conducted on the list of unique high priority-requirements corresponding to various stakeholders, by the SMEs, to identify whether these requirements were Gaps. 
Of the 56  gaps identified, 27 were found to pertain to both eVTOL and existing helicopter operations . Consequently, the latter highlighted a key area for the Regulator's attention. 
The final set of 56 identified gaps highlights specific, high-impact opportunities to strengthen regulatory frameworks across several key areas, including:
\begin{itemize}
    \item Organizational performance. 
    \item Process evaluation and improvement.
    \item Assessment criteria.
    \item Certification acknowledgment and confirmation.
    \item Training process for individuals involved in the regulatory process.
    \item Collision and Energy Management.
    \item Automation and Simulation.    
\end{itemize}

TABLE \ref{tab:TABLE 4} presents various stakeholder-specific gaps identified through the gap analysis, along with the corresponding recommendation type-whether regulatory, policy-related, or procedural. A comprehensive list of all the identified
gaps, linked to various stakeholders can be found in \cite{caa_stpa_evtol}. 
\par Addressing these identified gaps can significantly enhance the safety of eVTOL aircraft deployment and operations. The Regulator will need to incorporate the findings from this analysis into its broader regulatory programme. Several outcomes from the STPA have direct implications for the existing procedures, policies, and regulations. Where regulatory updates are necessary, the Rulemaking team would need to initiate the process of translating these findings into formal aviation legislations.

\begin{table}[hbt!]
\caption{\label{tab:TABLE 4} Some of the Gaps linked to various Stakeholders}
\centering
\begin{tabular}{p{0.2\linewidth}  p{0.7\linewidth}p{0.15\linewidth}} 
\hline
 \\
 \textbf{Stakeholder } & \textbf{Gap}  & \textbf{Recommendation Type}\\
 \hline
\textbf{Regulator(UK CAA)} & \textbf{UCA(Ph0.1)-16.1.1-RQ3:} Performance review of the relevant team issuing NOTAM within Regulator shall be conducted periodically to ensure that the team operates properly and safely. & Procedures \\
\textbf{Regulator(UK CAA)} & \textbf{UCA(Ph0.1)-24.2.1-RQ1:} The regulator shall train their staff adequately to ensure that the the supplementary documents regarding categorization of eVTOL, are reviewed properly. &  Procedures \\
 \textbf{Regulator(UK CAA)} & \textbf{UCA(Ph0.1)-14.5.1-RQ3:} The tasks related to processing the Temporary Airspace Structure within the Regulator should undergo routine review and be re-prioritized as necessary to guarantee that safety-critical tasks are prioritized above all others. &  Procedures \\
 \textbf{Regulator(UK CAA)} & \textbf{UCA(Ph0.1)-24.2.1-RQ3:} The assessment criteria for Categorisation
(CAT/NCC/NCO/SPO) shall be clearly presented to the applicant and shall
be consistent both internally within the Regulator and externally with the
applicant. &  Procedures \\
 \textbf{Air Navigation Service Provider(NATS)} & \textbf{UCA(Ph1)-21.1.1-RQ2:} The Meteorological conditions must be provided over a periodicity so that information is of a sufficient accuracy and available to NATS. & Procedures  \\
 \textbf{Air Navigation Service Provider(NATS)} & \textbf{UCA(Ph1)-18.2.1-RQ9:} There must be a mechanism for NATS to monitor and issue alerts when the performance (position, altitude, airspeed) of aircraft in flow is not within the expected values.  & Procedures  \\
 \textbf{Air Navigation Service Provider(NATS)} & \textbf{UCA(Ph1)-22.5.1-RQ2:} NATS shall ensure that modified flight plans, or new clearances are updated within \textit{to be defined} sec and managed centrally for all aircraft operating in a particular sector.   & Procedures  \\
  \textbf{eVTOL Operator}& \textbf{UCA(Ph0.1)-50.2.1-RQ8:} Process review of the relevant team issuing ‘Aircraft Release to service’ within the eVTOL Operator shall be conducted periodically to ensure that the team operates properly and safely. & Regulations\\
  \textbf{Vertiport Operator} & \textbf{UCA(Ph3)- 13.5.1-RQ.1:} Ground services must use advanced real-time sensors to ensure provision of continuous Feedback on landing conditions.  & Regulations\\
  \textbf{Vertiport Operator} & \textbf{UCA(Ph2)- 6.3.1-RQ.3:} Licensed Aerodrome shall utilise simulations to test the algorithm's efficiency to avoid unnecessary ‘Holds'.  & Procedures \\
 \textbf{Vertiport Operator} & \textbf{UCA(Ph0.2)-33.7.2-RQ6:} Local Authority shall ensure that the proposed ‘Number of Movements   \& Operational Hours’ are properly communicated with Temporary Aerodrome Management. & Policy \\
 \textbf{Vertiport Operator} & \textbf{UCA(Ph2)- 6.1.1-RQ.5:} Feedback to ‘Hold outside Restricted Area (Temporary) – 
RA(T)' regarding capacity status shall utilize multiple channels to ensure redundancy in communication pathways.  & Regulations \\
\hline
\end{tabular}
\end{table}

\section{Discussion}

\par Technology is evolving rapidly and today's systems have grown to be extensive, intricate, and increasingly automated. In evaluating potential failure causes, it is essential to consider not only the technical aspects but the entire socio-technical framework. Employees often deal with varied and at times conflicting information, swiftly changing scenarios, and significant workloads. A positive safety climate within an organization often results in fewer violations of safety protocols. STPA, which is based on a hierarchical safety control structure, is suitable for application to complex socio-technical systems, as it incorporates human, organizational, and procedural elements within the control framework. STPA for eVTOL identified several CF related to workload of the operators/controllers and degradation of internal processes over time. 

For example, one of the CF for UCA(Ph0.1)-28.2.1 in  TABLE \ref{tab:TABLE 2}) was ‘Regulator is unable to correctly process "Vertiport / Aerodrome Licence" due to the degradations of the internal processes over time (e.g., overloaded tasks, flawed process)'. A requirement was proposed to address this - ‘Process reviews for issuance of "Vertiport / Aerodrome Licence" and Performance reviews of the responsible team members shall be conducted periodically to ensure that the team operates efficiently and safely'.

One of the key values of STPA is that it provides a structured control-system–based framework for identifying Unsafe Control Actions and causal scenarios. This framework helps direct attention to all components and information pathways that may affect safety. As the analysis is based on the control structure and examines whether each control loop is a closed one, missing control inputs/ feedback signals can be easily identified. These missing input signals can then be incorporated into system design to improve it. For example, when analyzing the control loop for the control action UCA(Ph1)- 18.2.1 in TABLE \ref{tab:TABLE 2}, it becomes obvious that there is no means for the NATS Flight controllers to be aware of any deviations from flight plan of other aircraft and this could lead to incorrect issuance of Onward Clearance. UCA(Ph1)-18.2.1-RQ9 ( TABLE \ref{tab:TABLE 3}) was proposed to tackle this control loop flaw. 

 Traditional safety analysis techniques such as FTA and FMEA, place humans outside the system boundaries and consider them primarily as mitigators of hazardous physical system failures. These analysis techniques assume that humans are more perceptive and flexible in performance than machines and hence during adverse situations, human operators are expected to be able to deal with complex situations and system interactions. Although some methods like HAZOP consider human errors, they mostly consider only the operator errors and not the human factors in all levels of the hierarchy of a system. Human error is often viewed as random or probabilistic by these methods. However, human error—and behavior in general is shaped by the surrounding context and the design of the system \cite{henriksen2008understanding}. STPA integrates humans into the analysis just like any other component and offers support to understand why a human would make a mistake, thus leading to an accident. STPA entails that analysts evaluate how the design of a system can influence human errors, which may result in hazards, and subsequently design countermeasures to address these. STPA of eVTOL Operations identified operator mental‐model gaps linked to inadequate training, procedures, or supervision (Management flaws). For instance, various causes were associated with the misinterpretation of documents or information, and recommendations for sufficient training were suggested to prevent or mitigate these issues (see Examples: UCA(Ph2)-7.1.3-RQ.3 in TABLE \ref{tab:TABLE 3} and UCA(Ph0.1)-24.2.1-RQ1 in TABLE \ref{tab:TABLE 4}). To support human operators, particularly during times of high cognitive demand, recommendations were made to incorporate automation and simulation capabilities (UCA(Ph1)-18.2.1-RQ9 and UCA(Ph2)-6.3.1-RQ.3 in TABLE \ref{tab:TABLE 4}).

STPA provides guidance on identifying control actions using specific keywords – ‘too early’, ‘too late’, ‘stopped too soon’ and ‘applied too long’ to incorporate temporal factors. Time delays are crucial in designing control algorithms. If timing delays such as between issuing a command and the time the state change happens are not adequately considered in the control algorithms, accidents can happen. Several CFs related to delayed or inadequate feedback resulting in control actions being issued at the wrong time or for incorrect duration, were identified and requirements were proposed to address them (e.g.,  UCA(Ph1)-21.1.1-RQ2 and UCA(Ph3)- 13.5.1-RQ.1 in TABLE \ref{tab:TABLE 4}).

Several detailed causal factors related to communication issues were also identified. For instance, one of the CF for UCA(Ph1)-18.2.1 in TABLE \ref{tab:TABLE 3} was that the ‘OnwardClearance' correctly issued by the NATS (LHR RADAR) controller was incorrectly received by the Commander due to the signal interference or due to jamming or corruption of the signal. The analysis further underscored the necessity to enhance communication among various system components such as Vertiports/Aerodromes, the eVTOL Operators, and the Regulator to effectively manage various applications and approvals (e.g., UCA(Ph0.2)-33.7.2-RQ6 in TABLE \ref{tab:TABLE 4}). The analysis also highlighted the need to ensure coordination between system components for control instructions to be issued in a timely manner to ensure the safe operation of eVTOLs (e.g., UCA(Ph1)-22.5.1-RQ2 in TABLE \ref{tab:TABLE 4}). Similar to the traditional safety analysis techniques, the STPA of eVTOL operations also pinpointed the need for redundancy — for instance, in communication channels (see UCA(Ph2)-6.1.1-RQ.5 in TABLE \ref{tab:TABLE 4}).

This section outlines some of the practical challenges encountered and key lessons learned from this study. Some of the reviews of the STPA results (STPA Step-3 \& Step-4) were performed by a single individual representing a stakeholder. Hence, the results reflected the viewpoint of that individual. Given more time, a broader range of organizations and individuals could have been incorporated, to provide a more holistic view. For the prioritization of UCAs, it was observed that there were considerable differences in values assigned to different EJ factors for the same UCAs, by different individuals. Performing the EJ assessment for UCAs as a group activity for each stakeholder, was one way to address this. The Monte Carlo simulation (MCS) being a probabilistic model that can include an element of uncertainty or randomness in its prediction, another approach to tackle the subjective nature of EJ assessment was the application of MCS to reduce the uncertainties induced by differences in EJ factor values assigned for the same UCAs, by different individuals \cite{rose2023monte}. The authors adopted the latter and by combining inputs from STPA analysis with expert judgment and verifying these through MCS to reduce uncertainty, significant data was generated to assess risks and rank the UCAs effectively \cite{elbadaoui2025structured}. Handling and computing all the rankings for prioritizing the STPA results proved to be time-intensive. To enhance efficiency and accuracy, an automation tool was developed to streamline the prioritization process.
While STPA Step-1 and Step-2 reviews were initially conducted as group sessions with representatives from all stakeholder organizations, this approach proved less effective for later phases. As a result, the findings were organized by stakeholder, and individual review sessions were held with each stakeholder, to evaluate the results of STPA Step-3 and Step-4. One significant limitation of safety analysis techniques, such as STPA, lies in their subjectivity, largely due to the reliance on the analyst. Implementing a formalized methodology could help reduce this subjectivity, particularly when multiple safety analysts conduct the analysis, thereby enhancing its quality and efficiency. We used  a template that included a list of causal factors aligned with the different UCA categories tailored to the type of controller —be it human or machine described in \cite{elizebeth2025hazard} for the identification of causal factors in STPA Step 4, which improved the efficiency (in terms of effort) as well as reduced inter-analyst variations in the analysis.

\par Revisiting our research objectives, the STPA evaluation of eVTOL operations has shown that the current airspace design and regulatory framework is insufficiently prepared to manage potentially unexpected behaviors safely. The 56 gaps identified by the STPA for eVTOL Operations represent targeted, high-impact opportunities for regulatory enhancement in various aspects including organizational performance and process review (e.g., UCA(Ph0.1)-16.1.1-RQ3 in TABLE \ref{tab:TABLE 4}), assessment criteria(eg.,UCA(Ph0.1)-24.2.1-RQ3 in TABLE \ref{tab:TABLE 4}), acknowledgment and confirmation for certifications, and training process for individuals involved in the regulatory process (e.g.,UCA(Ph0.1)-24.2.1-RQ1 in TABLE \ref{tab:TABLE 4}). Gaps linked to collision and energy management (eg., UCA(Ph1)-18.2.1-RQ9 in TABLE \ref{tab:TABLE 4}), automation and simulation tools (eg., UCA(Ph2)- 6.3.1-RQ.3 in TABLE \ref{tab:TABLE 4}), and process improvement (eg., UCA(Ph0.1)-50.2.1-RQ8 in TABLE \ref{tab:TABLE 4}) were also identified. These gaps could potentially contribute to safer and more reliable deployment and operation of eVTOL aircraft in the future. Although not detailed in this paper, the concept for the prioritization of STPA results( Step-3 \& Step-4) \cite{ElBadaoui2025Prioritisation}\cite{chen2025scalable} which was developed as part of this work,  makes a significant, novel contribution towards objectively managing and identifying the results that require urgent
intervention and mitigation, to prevent catastrophic losses.
\par The STPA methodology played an important role in identifying potential safety and regulatory gaps by analyzing interactions within the system, based on the control structures. The findings from this analysis will serve as a foundation for future improvements, helping to refine the regulatory framework, policies and procedures to ensure they remain comprehensive, effective, and cater to the evolving aviation ecosystem.
From an industry standpoint, these insights can support the creation of Safety Management Systems for eVTOL stakeholders, which are essential for complying with the national Air Operations regulations. The Regulator is set to evaluate the integration of this study's findings, documented in a report \cite{caa_stpa_evtol}, into their regulatory work programme. The gaps identified through the STPA for eVTOL operations must be translated into actionable and practical measures for each relevant stakeholder. 
\section{Conclusion and Future Work}

This paper presents the application of STPA to assess the emergent behaviors and risks associated with the introduction of a novel technology - eVTOL into the current air navigation system. To help rank and prioritize the large number of results, a new prioritization method— an extension of the STPA was applied, which enabled the identification of a more targeted set of 124 high -priority requirements spanning multiple stakeholders. Upon comparison of the 124 requirements against existing aviation regulations, 56 were labeled as "gaps," meaning they were not covered by the current aviation regulations, policies and procedures . Among these identified gaps, 27 were noted to have an impact on both upcoming eVTOL operations as well as existing helicopter operations, emphasizing the need for their urgent resolution. 
\par The requirements pinpointed as "gaps" in the analysis must be transformed into actionable steps, tailored for each relevant stakeholder, by the Regulator. This should take the form of updated regulations and policies, operational directives, and practical guidance.This case study  has demonstrated that STPA is capable of effectively analysing the advanced features of the next generation aviation technologies and the complexity of the proposed operational improvements. The findings from this study demonstrate that STPA offers a thorough framework to identify shortcomings in existing regulations, policies, and procedures. This approach helps to establish a robust safety management system that proactively addresses risks and evolves in response to the challenges posed by new technologies.
\par As part of future work, a significant number of other lower priority requirements identified by the analysis will also need to be analyzed. Although the STPA analysis of eVTOL operations using the high-abstraction (Level 1) control structure has yielded valuable insights, future work could consider conducting an in-depth analysis using a more detailed (Level 2) control structure. This research could be extended to assess the potential risks associated with involving a wider range of eVTOL system stakeholders, like the National Power Network Operator and the local emergency authorities. It could also explore the consequences of automating air traffic management and integrating autonomous functionalities in eVTOLs. Considering the very limited research in this area, a comprehensive STPA analysis for the safety assessment of eVTOL aircraft is also worth investigating in the future.

\section{Acknowledgement}
The work presented in this paper was funded by the UK Department for Transport, delivered by the UK Civil Aviation Authority - GFA 3549. The authors would also like to thank the WMG center of HVM Catapult and WMG, University of Warwick, UK, for providing the necessary infrastructure for conducting this study. WMG hosts one of the seven centers that together comprise the High-Value Manufacturing Catapult in the UK.

\section{Acronym List}
    
    \begin{description}[leftmargin=*, widest=DCCHTM]
    \item[AAM]
    Advanced Air Mobility
        \item[CA]
        Control Action
        \item[CAA] 
        Civil Aviation Authority
        \item[CF]
        Commercial Air Transport
        \item[CAT] 
        Causal Factor
        \item[CIF]
        Controller Impact Factor
         \item[EASA]
         European Union Aviation Safety Agency
        \item[EJ]
        Expert Judgement
        \item[ETA] 
        Event Tree Analysis
        \item[eVTOL]
       electric vertical take-off and landing 
        \item[FAA]
        Federal Aviation Administration
        \item[FMEA]
        Failure Mode and Effects Analysis
        \item[FTA] 
        Fault Tree Analysis
        \item[HAZOP] 
        Hazard and Operability study
        \item[ICAO] 
        International Civil Aviation Organization
        \item[MCS] 
        Monte Carlo Simulation
         \item[NASA]
        National Aeronautics and Space Administration
        \item[NCC] 
        Non-commercial operations with complex motor-powered
aircraft
\item[NCO] 
Non-Commercial Operations with other than complex motor-
powered aircraft
        \item[NOTAM]
        Notice to Aviation
        \item[PED]
        Portable electronic devices
        \item[PMS] 
        Pre-mitigation Severity
        \item[RA(T)]
        Restricted Area (Temporary)
        \item[RF] 
        Radio Frequency
        \item[SME] 
        Subject Matter Expert
        \item[SPO] 
        Specialised Operations
        \item[STAMP] 
        System-Theoretic Accident Model and Processes
        \item[STPA] 
        Systems- Theoretic Process Analysis
        \item[UCA] 
        Unsafe Control Action
         \item[UAM]
        Urban Air Mobility
        \item[UAV]
        Unmanned Air Vehicles 
        \item[VFR] 
        Visual Flight Rules
        \item[VMC] 
        Visual Meteorological Conditions
       
    \end{description}

\bibliography{references}

\end{document}